\documentstyle[amssymb,aps]{revtex}

\begin{document}
\author{Remo Garattini}
\address{Facolt\`a di Ingegneria, Universit\`a degli Studi di Bergamo,\\
Viale Marconi, 5, 24044 Dalmine (Bergamo) Italy\\
e-mail: Garattini@mi.infn.it}
\title{Entropy from the foam II}
\maketitle

\begin{abstract}
A simple model of spacetime foam, made by two different types of wormholes
in a semiclassical approximation, is taken under examination: one type is a
collection of $N_{w}$ Schwarzschild wormholes, while the other one is made
by Schwarzschild-Anti-de Sitter wormholes. The area quantization related to
the entropy via the Bekenstein-Hawking formula hints a possible selection
between the two configurations. Application to the charged black hole are
discussed.
\end{abstract}

\section{Introduction}

The area-entropy relation, concerning black holes, has been proposed by J.
Bekenstein\cite{J.Bekenstein} in the early seventies. Despite of its simple
expression, it is still a central point of research in understanding the
connection between General Relativity and Quantum Mechanics. In natural
units one finds that the area-entropy becomes
\begin{equation}
S=\frac{A}{4G}=\frac{A}{4l_{p}^{2}},
\end{equation}%
where $A$ is the area of the event horizon, $G=l_{p}^{2}$ is the
gravitational constant. From one side, we have a statistical mechanics
problem and from the other side we have a pure geometrical problem. This is
another aspect of Einstein's equations relating geometry and dynamics.
However, the problem of understanding which kind of dynamics can give such a
simple result has yet to come. One progress has been made in terms of string
theory, where the entropy of a zero temperature black hole (extreme) has
been calculated and shown to be identical to the Bekenstein-Hawking formula
for the thermodynamical entropy\cite{SV}. This means that a count of states
of the black holes in terms of string states reflects the quantum nature of
a black hole. Another approach comes from the application of the Cardy's
formula\cite{Cardy} in conformal field theory\cite{Carlip,Strominger}
without invoking the string theory framework. Therefore a further
investigation from other points of view is as important as the string theory
approach. Having in mind the Bekenstein's proposal
\begin{equation}
a_{n}=\alpha l_{p}^{2}\left( n+\eta \right) \text{\qquad }\eta >-1\text{%
\qquad }n=1,2,\ldots ,
\end{equation}%
describing the quantization of the area for nonextremal black holes, in
Refs. \cite{RPLB2,RIJMPD,RNPBPS}, we have proposed a model made by $N_{w}$
wormholes, based on Wheeler's ideas of a foamy space-time\cite{Wheeler}. In
those papers, we have quantized the entropy of a Schwarzschild black hole
assuming the validity of the area-entropy relation. The area quantization
induced by the underlying foam background, whose quanta can be identified
with wormholes of Planckian size, has immediately led to the quantization of
the Schwarzschild black hole mass and of the cosmological constant. In this
paper we wish to generalize the results obtained in\ Refs.\cite%
{RPLB2,RIJMPD,RNPBPS} by looking at a foamy space composed by
Schwarzschild-Anti-de Sitter wormholes (S-AdS). A selection of the different
foamy constituents is suggested in terms of level spacings related to the
Hawking radiation. An application to a Reissner-Nordstr\"{o}m (RN) black
hole is also given. The rest of the paper is structured as follows, in
section \ref{p2} we briefly recall the results reported in Refs. \cite%
{RIJMPD,RCQG1} for the Schwarzschild and S-AdS wormholes respectively, in
section \ref{p3} we compute the area spectrum for both foam configurations,
in section \ref{p4} the area quantization is applied to the RN\ black hole.
We summarize and conclude in section \ref{p5}. Units in which $\hbar =c=k=1$
are used throughout the paper.

\section{Space-time foam: Schwarzschild or Schwarzschild-Anti-de Sitter
wormholes?}

\label{p2}We consider a complete manifold ${\cal M}$, divided in two wedges $%
{\cal M}_{+}$ and ${\cal M}_{-}$ by a bifurcation two-surface $S_{0}$,
located in the right and left sectors of a Kruskal diagram. We consider a
constant time hypersurface $\Sigma $ crossing $S_{0}$, representing an
Einstein-Rosen bridge with wormhole topology $S^{2}\times R^{1}$such that $%
\Sigma =\Sigma _{+}$ $\cup $ $\Sigma _{-}$. The line element we will
consider is
\begin{equation}
ds^{2}=-N^{2}\left( r\right) dt^{2}+\frac{dr^{2}}{1-\frac{b\left( r\right) }{%
r}}+r^{2}\left( d\theta ^{2}+\sin ^{2}\theta d\phi ^{2}\right) ,  \label{a1}
\end{equation}%
where $N\left( r\right) $ is the lapse function and $b\left( r\right) $ is
the shape function such that
\begin{equation}
b\left( r\right) =\left\{
\begin{array}{cc}
2MG & Schwarzschild \\
2MG-\frac{r^{3}}{b_{AdS}^{2}} & S-AdS%
\end{array}%
\right. .
\end{equation}%
$M$ is the wormhole mass, while $b_{AdS}^{2}=-3/\Lambda _{AdS}$ and $\Lambda
_{AdS}$ is the negative cosmological constant. Note that the throat is
located at $r_{h}=2MG$ in the Schwarzschild case, while for the S-AdS case
we get
\begin{equation}
1-\frac{2MG}{r_{h}}+\frac{r_{h}^{2}}{b_{AdS}^{2}}=0,  \label{a1b}
\end{equation}%
where we have implicitly defined $r_{h}$ in terms of $M$ and $b_{AdS}$. The
physical Hamiltonian defined on $\Sigma $ assumes the form
\begin{equation}
H_{P}=H_{\Sigma }+H_{\partial \Sigma },
\end{equation}%
where
\begin{equation}
H_{\Sigma }=\frac{1}{16\pi G}\int_{\Sigma }d^{3}x\left( N{\cal H}+N_{i}{\cal %
H}^{i}\right)
\end{equation}%
and
\begin{equation}
H_{\partial \Sigma }=\frac{1}{8\pi G}\int_{S_{+}}d^{2}x\sqrt{\sigma }\left(
k-k^{0}\right) -\frac{1}{8\pi G}\int_{S_{-}}d^{2}x\sqrt{\sigma }\left(
k-k^{0}\right) .
\end{equation}%
The volume term $H_{\Sigma }$ contains two constraints
\begin{equation}
\left\{
\begin{array}{l}
{\cal H}=G_{ijkl}\pi ^{ij}\pi ^{kl}\left( \frac{16\pi G}{\sqrt{g}}\right)
-\left( \frac{\sqrt{g}}{16\pi G}\right) \left( R^{\left( 3\right) }+\frac{6}{%
b_{AdS}^{2}}\right) =0 \\
{\cal H}^{i}=-2\pi _{|j}^{ij}=0%
\end{array}%
\right. ,  \label{a1a}
\end{equation}%
both satisfied by the Schwarzschild and flat metric respectively when $%
\Lambda _{AdS}=0$ $\left( b_{AdS}\rightarrow \infty \right) $ and satisfied
by the S-AdS and AdS metric when $b_{AdS}<\infty $. The supermetric is
defined as $G_{ijkl}=\frac{1}{2}\left(
g_{ik}g_{jl}+g_{il}g_{jk}-g_{ij}g_{kl}\right) $ and $R^{\left( 3\right) }$
denotes the scalar curvature of the surface $\Sigma $. The boundary term $%
H_{\partial \Sigma }$ (quasilocal energy\cite{BrownYork,FroMar}) is defined
by means of a subtraction procedure whose purpose is the elimination of the
asymptotic divergence, i.e. $r\rightarrow +\infty $ coming from the
curvature change\cite{HawHor}. In the present case, the Schwarzschild metric
asymptotically tends to the flat metric, which naturally defines the
reference space. The same happens for the S-AdS and the AdS spaces. In Refs. %
\cite{RCQG1,RIJMPA}, we have shown that $H_{\partial \Sigma }=0$ provided
that the boundary condition be symmetric with respect to the bifurcation
surface\footnote{%
See also Ref.\cite{FroMar}.}. Thus the Hamiltonian contribution comes from
the off-shell volume term. Nevertheless the form of $H$ is not satisfying,
because the subtraction procedure appears only at the boundary level.
However to be consistent with such a procedure, we extend the subtraction
even at the volume term. Thus the final physical Hamiltonian will be of the
form
\begin{equation}
\Delta H_{P}=H^{wormhole}-H^{No-wormhole}
\end{equation}%
and the change in energy can be computed with
\begin{equation}
\Delta E=\frac{\left\langle \Psi \left| H^{wormhole}-H^{No-wormhole}\right|
\Psi \right\rangle }{\left\langle \Psi |\Psi \right\rangle },  \label{a2}
\end{equation}%
by means of a variational approach, where the WKB functionals are
substituted with trial wave functionals. This quantity is the natural
extension to the volume term of the subtraction procedure for boundary terms
and it is interpreted as the Casimir energy related to vacuum fluctuations.
In practice, we consider perturbations at $\Sigma $ of the type
\begin{equation}
g_{ij}=\bar{g}_{ij}+h_{ij},
\end{equation}%
where $\bar{g}_{ij}$ is the spatial part of the Schwarzschild, Flat, S-AdS
and AdS background in a WKB approximation. By restricting our attention to
the graviton sector of the Hamiltonian approximated to second order,
hereafter referred to as $H_{|2}$, we define
\begin{equation}
E_{|2}=\frac{\left\langle \Psi ^{\perp }\left| H_{|2}\right| \Psi ^{\perp
}\right\rangle }{\left\langle \Psi ^{\perp }|\Psi ^{\perp }\right\rangle },
\end{equation}%
where
\begin{equation}
\Psi ^{\perp }=\Psi \left[ h_{ij}^{\perp }\right] ={\cal N}\exp \left\{ -%
\frac{1}{4}\left[ \left\langle \left( g-\bar{g}\right) K^{-1}\left( g-\bar{g}%
\right) \right\rangle _{x,y}^{\perp }\right] \right\} .
\end{equation}%
After having functionally integrated $H_{|2}$, we get
\begin{equation}
E_{|2}=\frac{1}{4}\int_{\Sigma }d^{3}x\sqrt{g}G^{ijkl}\left[ \left( 16\pi
G\right) K^{-1\bot }\left( x,x\right) _{ijkl}+\left( 16\pi G\right)
^{-1}\left( \triangle _{2}\right) _{j}^{a}K^{\bot }\left( x,x\right) _{iakl}%
\right] .
\end{equation}%
Thus Eq.$\left( \ref{a2}\right) $ becomes
\begin{equation}
\Delta E_{|2}=\frac{\left\langle \Psi ^{\perp }\left|
H_{|2}^{wormhole}\right| \Psi ^{\perp }\right\rangle }{\left\langle \Psi
^{\perp }|\Psi ^{\perp }\right\rangle }-\frac{\left\langle \Psi ^{\perp
}\left| H_{|2}^{No-wormhole}\right| \Psi ^{\perp }\right\rangle }{%
\left\langle \Psi ^{\perp }|\Psi ^{\perp }\right\rangle }.  \label{a2a}
\end{equation}%
The propagator $K^{\bot }\left( x,x\right) _{iakl}$ can be represented as
\begin{equation}
K^{\bot }\left( \overrightarrow{x},\overrightarrow{y}\right)
_{iakl}:=\sum_{N}\frac{h_{ia}^{\bot }\left( \overrightarrow{x}\right)
h_{kl}^{\bot }\left( \overrightarrow{y}\right) }{2\lambda _{N}\left(
p\right) },
\end{equation}%
where $h_{ia}^{\bot }\left( \overrightarrow{x}\right) $ are the
eigenfunctions of
\begin{equation}
\left( \triangle _{2}\right) _{j}^{a}:=-\triangle \delta _{j}^{a}+2V_{j}^{a}.
\end{equation}%
This is the Lichnerowicz operator projected on $\Sigma $ acting on traceless
transverse quantum fluctuations and $\lambda _{N}\left( p\right) $ are
variational parameters. $\triangle $ is the Laplacian in curved space
\begin{equation}
\triangle =\frac{1}{\sqrt{g}}\partial _{i}\left( \sqrt{g}g^{ij}\partial
_{j}\right)
\end{equation}%
and $V_{j\text{ }}^{a}$ is a mixed tensor containing the mixed Ricci tensor
whose components are:%
\begin{equation}
V_{j}^{a}=\left\{ -\frac{2MG}{r^{3}},\frac{MG}{r^{3}},\frac{MG}{r^{3}}%
\right\}
\end{equation}%
for the Schwarzschild case and

\begin{equation}
V_{j}^{a}=\left\{ -\frac{2MG}{r^{3}}+\frac{1}{b_{AdS}^{2}},\frac{MG}{r^{3}}+%
\frac{1}{b_{AdS}^{2}},\frac{MG}{r^{3}}+\frac{1}{b_{AdS}^{2}}\right\}
\end{equation}%
for the S-AdS case. The minimization with respect to $\lambda $ and the
introduction of a high energy cut-off $\Lambda $ give to Eq.$\left( \ref{a2a}%
\right) $ the following form
\begin{equation}
\Delta E\left( M,b_{AdS}\right) =\Delta E\left( M\right) =-\frac{V}{32\pi
^{2}}\left( \frac{3MG}{r_{0}^{3}}\right) ^{2}\ln \left( \frac{%
r_{0}^{3}\Lambda ^{2}}{3MG}\right) ,  \label{a3}
\end{equation}%
where $r_{0}>r_{h}$ and $V$ is the volume strictly localized to the wormhole
throat. Eq.$\left( \ref{a3}\right) $ is valid even when $b_{AdS}\rightarrow
\infty $. The minimum of $\Delta E\left( M\right) $ is located at $x_{2}=e^{-%
\frac{1}{2}}$, where $x=3MG/\left( r_{0}^{3}\Lambda ^{2}\right) $ and
\begin{equation}
\Delta E\left( x_{2}\right) =-\frac{V}{64\pi ^{2}}\frac{\Lambda ^{4}}{e}.
\end{equation}%
$\Delta E\left( x_{2}\right) $ shows a shift of the minimum away from the
expected one, namely $x_{1}=0$ corresponding to flat space and the AdS
space. The discrete spectrum contains exactly one mode. This gives the
energy an imaginary contribution, namely we have discovered an unstable mode %
\cite{RIJMPD,RCQG1}. This system can be stabilized if we consider $N_{w}$
wormholes in a semiclassical approximation and assume that there exists a
covering of $\Sigma $ such that $\Sigma =\bigcup\limits_{i=1}^{N_{w}}\Sigma
_{i}$, with $\Sigma _{i}\cap \Sigma _{j}=\emptyset $ when $i\neq j$. Each $%
\Sigma _{i}$ has topology $S^{2}\times R^{1}$ with boundaries $\partial
\Sigma _{i}^{\pm }$ with respect to each bifurcation surface. The boundary
is located at $R_{\pm }$ and it is reduced by a factor $N_{w}$, i.e. $R_{\pm
}\rightarrow R_{\pm }/N_{w}$. On each surface $\Sigma _{i}$, the boundary
Hamiltonian is
\begin{equation}
H_{\partial \Sigma _{i}^{\pm }}=\frac{1}{8\pi G}\int_{S_{i+}}d^{2}x\sqrt{%
\sigma }\left( k-k^{0}\right) -\frac{1}{8\pi G}\int_{S_{i-}}d^{2}x\sqrt{%
\sigma }\left( k-k^{0}\right) .
\end{equation}%
Note that $E_{\partial \Sigma _{i}^{\pm }}$ is zero for boundary conditions
symmetric with respect to {\it each} bifurcation surface $S_{0,i}$. We are
interested in a large number of wormholes, each of them contributing with a
term of the type $\left( \ref{a2}\right) $. The total semiclassical
hamiltonian is
\begin{equation}
H_{tot}^{N_{w}}=H^{1}+H^{2}+\ldots +H^{N_{w}}
\end{equation}%
and the total trial wave functional is the product of $N_{w}$ trial wave
functionals
\begin{equation}
\Psi _{tot}^{\perp }=\Psi _{1}^{\perp }\otimes \Psi _{2}^{\perp }\otimes
\ldots \ldots \Psi _{N_{w}}^{\perp }={\cal N}\exp N_{w}\left\{ -\frac{1}{4}%
\left[ \left\langle \left( g-\bar{g}\right) K^{-1}\left( g-\bar{g}\right)
\right\rangle _{x,y}^{\perp }\right] \right\} .
\end{equation}%
By repeating the same steps of the single wormhole, one gets
\begin{equation}
\Delta E_{N_{w}}\left( x,\Lambda \right) =N_{w}\frac{V}{32\pi ^{2}}\Lambda
^{4}x^{2}\ln x,
\end{equation}%
where we have defined the usual scale variable $x=3MG/\left(
r_{0}^{3}\Lambda ^{2}\right) $. Then at one loop the cooperative effects of
wormholes behave as one {\it macroscopic single }field multiplied by $N_{w}$%
, but without the unstable mode. At the minimum, $\bar{x}=e^{-\frac{1}{2}}$
\begin{equation}
\Delta E_{N_{w}}\left( \bar{x}\right) =-N_{w}\frac{V}{64\pi ^{2}}\frac{%
\Lambda ^{4}}{e}  \label{a4}
\end{equation}%
valid in presence or in absence of $\Lambda _{AdS}$.

\section{Area spectrum and Entropy}

\label{p3}Here we briefly recall how the area quantization process of a
black hole is obtained by means of the foam model. The area is measured by
the quantity
\begin{equation}
A\left( S\right) =\int_{S}d^{2}x\sqrt{\sigma }.
\end{equation}%
$\sigma $ is the two-dimensional determinant coming from the induced metric $%
\sigma _{ab}$ on the boundary $S$. The evaluation of the mean value of the
area
\begin{equation}
A\left( S\right) =\frac{\left\langle \Psi _{F}\left| \hat{A}\right| \Psi
_{F}\right\rangle }{\left\langle \Psi _{F}|\Psi _{F}\right\rangle }=\frac{%
\left\langle \Psi _{F}\left| \widehat{\int_{S}d^{2}x\sqrt{\sigma }}\right|
\Psi _{F}\right\rangle }{\left\langle \Psi _{F}|\Psi _{F}\right\rangle },
\end{equation}%
is computed on the following state
\begin{equation}
\left| \Psi _{F}\right\rangle =\Psi _{1}^{\perp }\otimes \Psi _{2}^{\perp
}\otimes \ldots \ldots \Psi _{N_{w}}^{\perp }.
\end{equation}%
Each single wormhole contributes with the quantity
\begin{equation}
A\left( S\right) =\frac{\left\langle \Psi _{F}\left| \hat{A}\right| \Psi
_{F}\right\rangle }{\left\langle \Psi _{F}|\Psi _{F}\right\rangle }=4\pi
\bar{r}^{2}.  \label{p31}
\end{equation}%
Suppose to consider the mean value of the area $A$ computed on a given {\it %
macroscopic} fixed radius $R$. On the basis of our foam model, we obtain $%
A=\bigcup\limits_{i=1}^{N}A_{i}$, with $A_{i}\cap A_{j}=\emptyset $ when $%
i\neq j$. Thus
\begin{equation}
A=4\pi R^{2}=\sum\limits_{i=1}^{N}A_{i}=4\pi
l_{p}^{2}\sum\limits_{i=1}^{N}x_{i}^{2}=4\pi l_{p}^{2}N\overline{x^{2}}=4\pi
l_{p}^{2}N\alpha ,  \label{p32}
\end{equation}%
where a new scale $x_{i}=\bar{r}_{i}/l_{p}$ has been introduced. $\alpha $
represents how each single wormhole area is distributed with respect to the
black hole area. Comparing Eq.$\left( \ref{p32}\right) $ with the Bekenstein
area spectrum proposal, we have
\begin{equation}
4\pi l_{p}^{2}N\alpha =4l_{p}^{2}N\ln 2
\end{equation}%
and $\alpha $ is fixed to
\begin{equation}
\frac{\ln 2}{\pi }=\alpha .
\end{equation}%
The entropy is simply
\begin{equation}
S=\frac{A}{4l_{p}^{2}}=N\ln 2  \label{p33}
\end{equation}%
and for the Schwarzschild geometry we get
\begin{equation}
S=\frac{4\pi \left( 2MG\right) ^{2}}{4G}=4\pi M^{2}G=4\pi
M^{2}l_{p}^{2}=N\ln 2.  \label{p34}
\end{equation}%
It is immediate to see that
\begin{equation}
M=\frac{\sqrt{N}}{2l_{p}}\sqrt{\frac{\ln 2}{\pi }},  \label{p35}
\end{equation}%
namely the Schwarzschild black hole mass is {\it quantized} in terms of $%
l_{p}$ which is in agreement with the results presented in Refs.\cite%
{Ahluwalia,Hod,Makela,VazWit,Mazur,Kastrup,JGB}. This implies also that the
level spacing of the transition frequencies is
\begin{equation}
\omega _{0}=\Delta M=\left( 8\pi Ml_{p}^{2}\right) ^{-1}\ln 2.  \label{p36}
\end{equation}%
When we use Eq. (\ref{p33}) for the de Sitter geometry, we get
\begin{equation}
S=N\ln 2=\frac{3\pi }{l_{p}^{2}\Lambda _{c}}=\frac{A}{4l_{p}^{2}}=\frac{%
N4\pi l_{p}^{2}}{4l_{p}^{2}}=N\pi ,
\end{equation}%
that is
\begin{equation}
\frac{3\pi }{l_{p}^{2}N\ln 2}=\Lambda _{c}.  \label{p37}
\end{equation}%
An interesting aspect appears when we put numbers in Eq.$\left( \ref{p37}%
\right) $. When $N=1$, the foam system is highly unstable and the
cosmological constant assumes the value, in order of magnitude, of $\Lambda
_{c}\sim 10^{38}GeV^{2}$. However the system becomes stable when the whole
universe has been filled with wormholes of Planckian size and this leads to
the huge number $N=10^{122}$ corresponding to the value of $\Lambda _{c}\sim
10^{-84}GeV^{2}$ which is the order of magnitude of the cosmological
constant of the space in which we now live.

\subsection{S-AdS Space-Time Foam}

In Section \ref{p2}, we have briefly reported how a model of space-time foam
formed can be realized by S-AdS wormholes. Even if the computation has been
done for a single wormhole, the procedure of a large $N_{w}$ S-AdS wormhole
approach can be realized straightforwardly in analogy with the Schwarzschild
case\footnote{%
See Ref.\cite{RCQG1} for details.}. To consider a large $N_{w}$ approach to
space-time foam even with S-AdS wormholes, one has to consider the following
rescaling
\begin{equation}
\left\{
\begin{array}{c}
R_{\pm }\rightarrow R_{\pm }/N_{w} \\
l_{p}^{2}\rightarrow N_{w}l_{p}^{2} \\
\Lambda _{AdS}\rightarrow \Lambda _{AdS}/N_{w}^{2}%
\end{array}%
\right. ,
\end{equation}%
where $R_{\pm }$ are the boundaries related to a single wormhole. This
rescaling is a consequence of the boundary reduction related to the
semiclassical superposition of wormholes wave functionals leading to a
stable system. From Eq.$\left( \ref{a4}\right) $, we see that both
representations of the foam, i.e. S-AdS and Schwarzschild wormholes, have
the same energy contribution. Thus we need one more information to select
which representation seems to be the correct one. This information comes
exactly from the area-entropy quantization applied to a S-AdS black hole.
The procedure is simply a repetition of what has been done in the
Schwarzschild case but with S-AdS wormholes. Let us consider a S-AdS black
hole whose horizon is located at $r_{h}$. Then from Eq.$\left( \ref{a1b}%
\right) $, we write
\begin{equation}
M_{S-AdS}=\frac{r_{h}}{2l_{p}^{2}}\left( 1+\frac{r_{h}^{2}}{b_{AdS}^{2}}%
\right)  \label{sads1}
\end{equation}%
which tends to the Schwarzschild mass $M$ when $b_{AdS}\rightarrow \infty $.
The application of Eq.$\left( \ref{p32}\right) $ to the area of the horizon
gives
\begin{equation}
r_{h}=l_{p}\sqrt{N\alpha _{AdS}}.
\end{equation}%
$\alpha _{AdS}$ represents the size of each S-AdS wormhole inside the black
hole horizon. However Eq.$\left( \ref{sads1}\right) $ does not uniquely
define a black hole. To this purpose, we consider the value $r_{h,m}=b_{AdS}/%
\sqrt{3}$ obtained by minimizing the surface gravity with respect to the
horizon radius. Note that in the terminology of the black hole
thermodynamics $r_{h,m}$ corresponds to the unique black hole solution whose
temperature reaches its minimum. Thus $b_{AdS}=\sqrt{3}l_{p}\sqrt{N\alpha
_{AdS}}$ and
\begin{equation}
M_{AdS}=\frac{2\sqrt{\alpha _{AdS}N}}{3l_{p}}%
\mathrel{\mathop{\longrightarrow }\limits_{b_{AdS}^{2}\rightarrow \infty }}%
M_{S}=\frac{\sqrt{N}}{2l_{p}}\sqrt{\frac{\ln 2}{\pi }.}
\end{equation}%
This fixes $\alpha _{AdS}$ to
\begin{equation}
\alpha _{AdS}=\frac{9\ln 2}{16\pi }.  \label{sads2}
\end{equation}%
A straightforward consequence of Eq.$\left( \ref{sads2}\right) $is that the
transition frequencies of the emitted radiation of a black hole can have a
different spectrum. Indeed, if the foam is represented by Schwarzschild
wormholes, the level spacing observed is given by Eq.$\left( \ref{p36}%
\right) $ in both cases, i.e. the S-AdS and Schwarzschild black hole.
Nevertheless, if the foam is represented by S-AdS wormholes, the level
spacing of a Schwarzschild black hole is given by
\begin{equation}
\omega _{0}=\Delta M_{S}^{AdS}=\left( 8M_{S}l_{p}^{2}\right) ^{-1}\alpha
_{AdS}=\left( 8M_{S}l_{p}^{2}\right) ^{-1}\frac{9\ln 2}{16\pi }=\Delta M%
\frac{9}{16}
\end{equation}%
that it means that for a given Schwarzschild black hole of mass $M$, the
S-AdS foam representation gives smaller frequencies.

\section{The Reissner-Nordstr\"{o}m black hole}

\label{p4}In this section we would like to apply the foam covering to the
case of a black hole with a charge. A charged black hole is described by
\begin{equation}
ds^{2}=-f\left( r\right) dt^{2}+f\left( r\right) ^{-1}dr^{2}+r^{2}d\Omega
^{2},  \label{p41}
\end{equation}
with
\begin{equation}
f\left( r\right) =\left( 1-\frac{2MG}{r}+\frac{G\left(
Q_{e}^{2}+Q_{m}^{2}\right) }{r^{2}}\right) =\left( 1-\frac{2MG}{r}+\frac{%
Q^{2}}{r^{2}}\right) .  \label{p42}
\end{equation}
$Q_{e}$ and $Q_{m}$ are the electric and magnetic charge respectively. When $%
Q=0$ the metric describes the Schwarzschild metric, while $Q=M=0$ describe a
flat metric. For $Q\neq 0$, we can distinguish three different cases:

\begin{itemize}
\item[a)] $MG>Q$. In this case the gravitational potential $f\left( r\right)
$ admits two real distinct solutions located at
\begin{equation}
\left\{
\begin{array}{c}
r_{+}=MG+\sqrt{\left( MG\right) ^{2}-Q^{2}} \\
r_{-}=MG-\sqrt{\left( MG\right) ^{2}-Q^{2}}%
\end{array}
\right. ,
\end{equation}
with $f\left( r\right) >0$ for $r>r_{+}$ and $0<r<r_{-}$ . $r_{-}$ is a
Cauchy horizon and $r_{+}$ is an event horizon. For each root there is a
surface gravity defined by
\begin{equation}
\kappa _{\pm }=\lim\limits_{r\rightarrow r_{\pm }}\frac{1}{2}\left|
g_{00}^{\prime }\left( r\right) \right| ,
\end{equation}
whose values are
\begin{equation}
\left\{
\begin{array}{c}
\kappa _{+}=\left( r_{+}-r_{-}\right) /2r_{+}^{2} \\
\kappa _{-}=\left( r_{-}-r_{+}\right) /2r_{-}^{2}%
\end{array}
\right. .
\end{equation}
The {\it Hawking temperature} associated with the surface gravity of the
event horizon is
\begin{equation}
T_{H}=\frac{\kappa _{+}}{2\pi }.
\end{equation}

\item[b)] $MG=Q$. This is the extreme case. The gravitational potential $%
f\left( r\right) $ admits two real coincident solutions located at $%
r_{+}=r_{-}=r_{e}=MG$ and its form is $f\left( r\right) =\left(
1-MG/r\right) ^{2}$. Here we discover that $\kappa _{+}=\kappa _{-}=0$ and $%
T_{H}=0$.

\item[c)] $MG<Q$. In this case the gravitational potential $f\left( r\right)
$ admits two complex conjugate solutions located at
\begin{equation}
\left\{
\begin{array}{c}
r_{+,i}=MG+i\sqrt{Q^{2}-\left( MG\right) ^{2}} \\
r_{-,i}=MG-i\sqrt{Q^{2}-\left( MG\right) ^{2}}%
\end{array}
\right. ,
\end{equation}
respectively.
\end{itemize}

Cases a) and b) imply $Q=0$ when $M=0$. We will consider the application of
the area quantization to cases a) and b). The constant $\alpha $, describing
the ``{\it fine structure}'', will be left unspecified for the whole
computation. To have a matching with the Bekenstein proposal we express the
mass $M$ of the charged black hole in terms of the area of the event horizon
and its charge $Q$. This is easily done by invoking the
Christodoulou-Ruffini formula\cite{CR}
\begin{equation}
M=\left[ \frac{A}{16\pi G^{2}}\left( 1+\frac{4\pi Q^{2}}{A}\right) ^{2}%
\right] ^{\frac{1}{2}}  \label{p43}
\end{equation}
obtained by inverting
\begin{equation}
A=4\pi r_{+}^{2}=4\pi \left( MG+\sqrt{\left( MG\right) ^{2}-Q^{2}}\right)
^{2}.  \label{p44}
\end{equation}
We observe that if $Q=0$, then we recover the Schwarzschild case, while if $%
A=4\pi Q^{2}$, we have the extreme one. The application of Eq.$\left( \ref%
{p32}\right) $ to the area of the RN black hole gives
\begin{equation}
A=4\pi \alpha l_{p}^{2}N_{2},  \label{p45}
\end{equation}
where $N_{2}$ is the wormholes number used for the covering of the RN black
hole area. However, from Eq.$\left( \ref{p44}\right) $we obtain that
\begin{equation}
Q^{2}=\sqrt{N_{2}}\left( \sqrt{N_{1}}-\sqrt{N_{2}}\right) \alpha l_{p}^{2},
\label{p46}
\end{equation}
where we have used Eq.$\left( \ref{p35}\right) $. We immediately see that
from the above equation we have $N_{1}\geq N_{2}$, where the equality
corresponds to the vanishing charge. Moreover we choose $N_{1}$ and $N_{2}$
in such a way that $Q^{2}=\alpha l_{p}^{2}q$, $q=0,1,2,\ldots $. This means
that
\begin{equation}
\sqrt{N_{2}}\left( \sqrt{N_{1}}-\sqrt{N_{2}}\right) =q.  \label{p46a}
\end{equation}
Note that if we put Eq.$\left( \ref{p46}\right) $ into Eq.$\left( \ref{p43}%
\right) $, we obtain always the Schwarzschild case. This means that the
imposed condition of Eq.$\left( \ref{p46a}\right) $ reveals the physics of
the charged black hole. By solving with respect to $N_{2}$, we get
\begin{equation}
N_{2}=-q+\frac{N_{1}}{2}\pm \frac{\sqrt{N_{1}}\sqrt{N_{1}-4q}}{2}.
\label{p47}
\end{equation}
with $N_{1}\geq 4q$. When $q=0$, we recover the Schwarzschild case, namely $%
N_{2}=N_{1}$ which implies the plus choice in Eq.$\left( \ref{p47}\right) $%
\footnote{%
This is only Eq.$\left( \ref{p44}\right) $ written in terms of quantum
numbers.}. On the other hand, when we consider $N_{1}=4q$, we get $N_{2}=q$
corresponding to the extreme case. Inserting Eqs.$\left( \ref{p45}\right) $
and $\left( \ref{p46a}\right) $ into Eq.$\left( \ref{p43}\right) $ we get
\begin{equation}
M=\frac{\sqrt{\alpha }}{2l_{p}}\sqrt{N_{2}}\left( 1+\frac{q}{N_{2}}\right) .
\label{p48}
\end{equation}
It is interesting to see what happens on the level spacing when we consider
a transition in mass with a fixed charge\footnote{%
See also Refs.\cite{BDK,MR}}. From Eq.$\left( \ref{p44}\right) $ we get
\begin{equation}
\omega _{0}=\Delta M=\frac{\partial M}{\partial r_{+}}\frac{dr_{+}}{dN}%
\Delta N=\frac{\pi }{A}\left( r_{+}-r_{-}\right) \alpha  \label{p49}
\end{equation}
with $\Delta N=1$, which vanishes in the extreme limit\cite{BDK,VazWit1}. It
is also interesting to see that the wormhole model perfectly agrees with the
fourth postulate of Ref.\cite{Makela1}, asserting that:

{\it If }$E_{i}$ {\it and }$E_{f}$ {\it are the initial and the final
energies which may be extracted from the horizon, and }$n_{i}A_{0}$ {\it and
}$n_{f}A_{0}$ {\it are the corresponding horizon areas, then}
\begin{equation}
\int_{E_{i}}^{E_{f}}\frac{dE}{\kappa \left( E\right) }=\frac{\ln 2}{2\pi }%
\left( n_{f}-n_{i}\right) ,  \label{p410}
\end{equation}%
{\it where }$\kappa \left( E\right) $ {\it is the surface gravity of the
horizon.} Let us apply Eq.$\left( \ref{p410}\right) $ to the RN black hole.
The left hand side becomes
\begin{equation}
\int_{M_{i}}^{M_{f}}\frac{dM}{\kappa \left( M,Q\right) }=\int_{M_{i}}^{M_{f}}%
\frac{r_{+}^{2}dM}{\sqrt{\left( MG\right) ^{2}-Q^{2}}}=\left[ M^{2}G+M\sqrt{%
\left( MG\right) ^{2}-Q^{2}}\right] _{M_{i}}^{M_{f}}=\left[ Mr_{+}\right]
_{M_{i}}^{M_{f}}.  \label{p411}
\end{equation}%
We now use Eqs.$\left( \ref{p44},\ref{p48}\right) $ and Eq.$\left( \ref{p411}%
\right) $ becomes
\begin{equation}
\left[ \frac{\alpha N_{2}}{2}\left( 1+\frac{q}{N_{2}}\right) \right]
_{N_{i}}^{N_{f}}=\frac{\alpha }{2}\left( N_{2,f}-N_{2,i}\right) ,
\label{p412}
\end{equation}%
which is the right hand side of fourth postulate. Note that the value of $%
\alpha $ which matches with the postulate is given by the Schwarzschild foam
model. Indeed if the foam model is represented by the S-AdS wormholes we
have a factor $9/16$ that changes the postulate.

\section{Conclusions}

\label{p5}In this paper we have applied the model presented in Refs.\cite%
{RPLB2,RIJMPD,RNPBPS}\ to a larger types of black holes including the
negative cosmological constant and a charge. Even in this case, assuming the
validity of the Bekenstein-Hawking relation, the entropy has been ``{\it %
quantized}''. Precisely, it is the area that it has been quantized; this is
the effect of a space-time filled by a given integer number of disjoint
non-interacting wormholes. Nevertheless, we have two possibility of
reproducing a foamy space-time:

\begin{enumerate}
\item A foamy space-time made by Schwarzschild wormholes.

\item A foamy space-time made by Schwarzschild-Anti-de Sitter wormholes.
\end{enumerate}

This means that we need a selection mechanism. This is given exactly by a
possible observation of the spectrum of the Hawking radiation and in
particular from the level spacings of the black hole under examination. We
recall that the level spacings appear because the black hole mass has been
quantized in terms of wormholes. Note that this quantization procedure is in
agreement with the quantized area proposed heuristically by Bekenstein and
reproduced by different authors\cite%
{Ahluwalia,Hod,Makela,VazWit,Mazur,Kastrup,JGB}. The degeneracy factor $%
\left( \ln 2\right) $ is here interpreted as an effect of the invariance of
the orientation of the wormhole with respect to the black hole area,
erroneously interpreted in Ref.\cite{RPLB2} as an effect of having a model
of an ``{\it ideal Boltzmann gas''} of wormholes. The statistics of the foam
wave function has never been introduced, but however the logarithmic factor
appears due to the average distribution of the different radius sizes: it is
a pure geometrical effect, that in the General Relativity language means
dynamical effect. The covering procedure applied to the RN black hole
introduces a second quantum number related to the charge. Nevertheless this
number comes in because a different horizon with different forces has been
considered. This means that even the charge quantum number could be related
to a geometrical/dynamical effect\cite{Cavaglia}. Note that also the extreme
case can be easily considered. Concerning this point there exists an open
problem about the validity of the area-entropy relation\footnote{%
There are arguments coming from a combination between quantum, thermodynamic
and statistical arguments\cite{Hod1} leading to a failure of the
entropy-area relation.}. Indeed, there exist well known indications that in
the extreme RN black hole, the entropy is zero due to the strict relation
between entropy and topology\cite{HHR,GK,Teitelboim,LP}. This conclusion
comes from the following equation%
\begin{equation}
S=\chi \frac{A}{4G},
\end{equation}%
where $\chi $ is the Euler characteristic\footnote{%
In Ref.\cite{LP}, the formula proposed for the area-entropy relation has
been modified in%
\begin{equation}
S=\chi \frac{A}{8G}.
\end{equation}%
This formula has been further generalized in Ref.\cite{Wu}, where
the author
considers the extension to topological black holes.}. Since the value of $%
\chi $ is zero for the extreme RN, the entropy is zero too. How the foam can
approach this problem will be the subject of a future investigation.

\section{Acknowledgments}

I wish to thank J. Bekenstein, T. Jacobson, J. M\"{a}kel\"{a}, R. Mann and
C. Vaz for useful comments and discussions. In particular, I would like to
thank M. Cavagli\`{a} who has brought to my attention Refs.\cite%
{Carlip,Strominger}.

\end{document}